\documentclass[%
 reprint,
 amsmath,amssymb,
 aps,
]{revtex4-2}

\usepackage{graphicx}
\usepackage{dcolumn}
\usepackage{bm}
\usepackage{braket}
\usepackage{color}

\newcommand{\Lnorm}[1]{\left\lVert#1\right\rVert}

\begin{document}

\preprint{APS/123-QED}

\title{Non-Uniqueness of Non-Linear Optical Response}

\author{Gerard McCaul}%
 \email{gmccaul@tulane.edu}
 \affiliation{Tulane University, New Orleans, LA 70118, USA}
 
\author{Alexander F. King}
 \email{aking20@tulane.edu}
 \affiliation{Tulane University, New Orleans, LA 70118, USA}
 
\author{Denys I. Bondar}
 \email{dbondar@tulane.edu}
\affiliation{Tulane University, New Orleans, LA 70118, USA}

\date{\today}

\begin{abstract}
In recent years, non-linear optical phenomena have attracted much attention, with a particular focus on the engineering and exploitation of non-linear responses. Comparatively little study has however been devoted to the driving fields that generate these responses. In this work, we demonstrate that the relationship between a driving field and the optical response it induces is not-unique. Using a generic model for a strongly interacting system, we show that multiple candidate driving field exists, which will all generate the same response. Consequently, it is possible show that the optical response is not sufficient to determine the internal dynamics of the system, and that different solutions for the driving field will do different amounts of work on a system. This non-uniqueness phenomenon may in future be utilised to engineer internal system states without modifying its optical response.
\end{abstract}

\maketitle


\section{\label{sec:intro}Introduction}
A rather ubiquitous old saw is that each person is unique \cite{Varki2008}. By the same token, it is often assumed that physical systems share an analogous property, i.e. that their response to external driving will be uniquely determined by their material properties. This is unsurprising, as the vast majority of physics is studied in the regime of linear response. Any linear equation will only have one solution, and the material simply determines the constant of proportionality between driving field and response. In fact, such properties at equilibrium serve to \emph{define} a material. Many empirical laws, like Snell's law and Ohm's law are consequences of this linear approximation between an incoming electromagnetic field and the system's response \cite{jackson_classical_1999}. 

 It is important to remember however, that the linear response approximation is just that - an \emph{approximation}. While non-linear optical phenomena were predicted \cite{goppertmayer_uber_1931} before the advent of lasers, outside of a few exceptional cases light intensities were too low for non-linear effects to be observed experimentally \cite{lewis_reversible_1941,kerr_xl_1875}. In recent years however the study of strong field physics has progressed significantly, and a plethora of nonlinear effects have been observed. 
 
 Principal among these non-linear effects is  High Harmonic Generation (HHG) \cite{Ghimire2012, Ghimire2011a, Murakami2018, PhysRevB.103.035110,Corkum2007,RevModPhys.81.163,Li2020}, which has been considered as a potential source of extreme ultra violet and soft x-ray light \cite{silva_high-harmonic_2018,yang_high-harmonic_2019}. Additionally, the exploitation of non-linear optical effects has spawned the study of meta-materials to be able to engineer non-linear optical materials with specific properties \cite{lapine_colloquium_2014,sain_nonlinear_2019}. These have been used to produce novel optical effects such as cloaking and super-resolution \cite{leonhardt_focus_2008,pendry_negative_2000}, as well as the production of materials with heightened optical non-linearities \cite{krasnok_nonlinear_2018,kauranen_nonlinear_2012}. 
 
 A natural consequence of a non-linear responses, is that the uniqueness of the response to a driving field, is \emph{no longer guaranteed}. Consider that even the simplest model of centro-symmetric nonlinear optics, where the polarisation $Y(t)$ is described by $Y(t) = \chi^{(1)}E(t) + \chi^{(3)}E^3(t)$ \cite{boyd_nonlinear_2008}, will have (depending on the susceptibilities $\chi^{(1)},\chi^{(3)}$) up to 3 solutions for the applied field $E(t)$. In general, if the response is determined by a polynomial of degree $n$, there will be $n$ solutions to the applied field \cite{brown_complex_2009}. Unfortunately, for many systems an accurate description of the response as a high order polynomial in $E(t)$ is a task of formidable difficulty, meaning that outside of some simple systems, it is not possible to enumerate every solution. 
 
 This difficulty has led to non-uniqueness being a little-studied topic in optics, particularly given its focus on the application and exploitation of non-linear effects, rather than the fields which cause them. The phenomenon of non-uniqueness has however appeared sporadically in a number of topics ranging from recovering scattering potentials from reflection and transmission coefficients \cite{abraham_two_1981}, to the existence of dissimilar Hamiltonians in a Dirac field theory predicting identical probability currents \cite{arminjon_non-uniqueness_2011}. In the context of quantum metrology devices, the ability to uniquely determine an input driving field from some observable is of paramount importance, and is relied upon in (for instance), quantum non-demolition measurements of squeezed light in a gravitational wave detector \cite{ong_invertibility_1984,clark_quantum_1985}. An area where non-uniqueness is of particular interest is quantum control \cite{PRXQuantum.2.010101, Ong1984,Clark1985,PhysRevA.72.023416,PhysRevA.98.043429,PhysRevA.84.022326,Campos2017,doi:10.1063/1.1582847,doi:10.1063/1.477857,2010.13859}. As optimal quantum control made the manipulation of chemical reactions through light feasible \cite{peirce_optimal_1988,shi_optimal_1988,gross_inverse_1993}, it was observed that multiple driving fields could be used to generate the same expectation trajectory for a 3-level quantum system \cite{jha_multiple_2009}.
 
 In this paper, we generalise this observation, and demonstrate that response non-uniqueness is in fact a generic feature of driven quantum systems. Using recently developed many-body control procedures \cite{mccaul_controlling_2020,mccaul_driven_2020}, we derive the conditions under which non-uniqueness can occur in Sec.\ref{sec:Model}. Sec.\ref{sec:Numerics} then applies this framework, where sets of driving fields which all generate the same system response are calculated. The physical consequences of a particular choice of solution are also investigated, and it is shown that different driving fields will perform different amounts of work on the system, despite generating the same response. Finally, we close the paper with a discussion of the results in Sec.\ref{sec:Discussion}.

\section{Model \label{sec:Model}}
We begin with an exposition of response uniqueness in quantum dynamics, and the manner in which this may be violated. Consider first a (finite dimensional) quantum system with Hamiltonian $\hat{H}$ evolved via the Schr{\"o}dinger equation,
\begin{equation}
    i \frac{{\rm d}}{{\rm d t}} \ket{\psi} = \hat{H}\ket{\psi}. \label{eq:schrodeq}
\end{equation}
Given this is a linear dynamical equation \cite{Jordan_2009}, there is a \emph{prima facie} case that this evolution will be unique provided one uses a bounded Hamiltonian. The most direct route to establishing this is to show that $\hat{H}$ is \emph{Lipschitz continuous} (LC) over the space of states $\ket{\psi}$ \citep{geraldfolland2007}. This condition  then guarantees $\ket{\psi}$ has a unique solution depending on its initial value via the Picard-Lindel{\"o}f theorem  \citep{kentnagle2011}. 

An operator $\hat{A}$ is LC if for all $\ket{\psi}, \ket{\phi}$, it satisfies the following condition:
\begin{equation}
\Lnorm{\hat{A}\ket{\psi}- \hat{A}\ket{\phi}} \leq L_A \Lnorm{\ket{\psi}-\ket{\phi}}, \label{eq:TrackingLipschitz}
\end{equation}
where $\Lnorm{\cdot}$ is any submultiplicative norm (which for convenience and consistency with the Born rule we take to be the 2-norm), and  the Lipschitz constant $L_A$ is any finite number. For any bounded operator, this condition is automatically satisfied, with 
\begin{equation}
    L_A=\sup_{\braket{\psi|\psi}=1}\Lnorm{\hat{A}\ket{\psi}}.
\end{equation}

Clearly, any system evolving according to Eq.\eqref{eq:schrodeq} will be LC, and therefore both $\ket{\psi}$ and any expectations calculated from it will have a unique solution. For there to be a possibility of non-uniqueness, there must be some \emph{non-linearity} present in the dynamics. While the Schr{\"o}dinger equation itself is guaranteed to be linear, there are many examples of highly non-linear phenomena in driven systems. Prominent among these are optical effects, where the relationship between the driving field $E(t)$ and optical response $J(t)$ can exhibit extreme non-linearity. This is most clearly observed in the phenomenon of HHG, and can be exploited to force distinct systems to exhibit identical responses \cite{mccaul_driven_2020, mccaul_indistinguishability}, or perform efficient computations \cite{2104.06322}.

Here we demonstrate that this nonlinear input-output relationship is non-unique, and can be leveraged to derive a multiplicity of driving fields which all generate the same response $J(t)$. We begin by modelling a many-electron system interacting with a classical field via the dipole approximation \cite{mccaul_controlling_2020,2104.06322}. For a lattice system using fermionic annhilation operators $\hat{c}_{j\sigma}$, this leads to the Hamiltonian:
\begin{equation}
\hat{H}\text{\ensuremath{\left(t\right)}}=  -t_{0}\sum_{j,\sigma}\text{\ensuremath{\left({\rm e}^{-i\Phi\left(t\right)}\hat{c}_{j\sigma}^{\dagger}\hat{c}_{j+1\sigma}+{\rm h.c.}\right)}}  +\hat{U} \label{eq:Hamiltonian}
\end{equation}
where $t_0$ is the hopping parameter, $\Phi(t)=aA(t)$ is the Peierls phase, $a$ is the system lattice constant and $A(t)=\int^t_0 {\rm d}t^\prime E(t^\prime)$ is an electromagnetic vector potential. Finally $\hat{U}$ specifies the electron-electron interactions, and we stipulate only that it commutes with the number operator $\hat{n}_{j\sigma}=\hat{c}^\dagger_{j\sigma}\hat{c}_{j\sigma}$.

For a given phase $\Phi(t)$, this evolution will be unique, and hence have a unique response $J(t)$. This response is derived from a continuity equation for the electron density \cite{mccaul_indistinguishability}, which defines the current operator
\begin{equation}
\hat{J}=-iat_{0}\sum_{j,\sigma}\left({\rm e}^{-i\Phi\left(t\right)}\hat{c}_{j\sigma}^{\dagger}\hat{c}_{j+1\sigma}-{\rm h.c.}\right),\label{eq:currentoperator}
\end{equation}
with $J(t)=\braket{\psi|\hat{J}|\psi}$. 

While the trajectory of $\Phi(t)$ (and the initial state of the system) will always determine $J(t)$ uniquely, it is possible to show the converse is not true, even under the condition that $\Phi(t)$ be continuous.  That is, there are multiple trajectories of $\Phi(t)$ which produce distinct dynamics (i.e. excluding the trivial $2 \pi$ periodicity of $\Phi(t)$), but result in the \emph{same} response $J(t)$.  

To show this, we first express $\Phi(t)$ as a function of the current $J(t)$. This is most easily achieved by expressing the nearest neighbour expectation in a polar form:
\begin{equation}
\left\langle \psi (t) \left|\sum_{j,\sigma} \hat{c}_{j\sigma}^{\dagger}\hat{c}_{j+1\sigma}\right| \psi (t) \right\rangle =R\left(\psi\right){\rm e}^{i\theta\left(\psi\right)}. \label{neighbourexpectation}
\end{equation}
In both Eq.\eqref{neighbourexpectation} and later expressions, the  argument $\psi$ indicates that the expression is dependent on $\ket{\psi}\equiv \ket{\psi(t)}$. Eq.\eqref{neighbourexpectation} can be used in conjunction with Eq.\eqref{eq:currentoperator} to yield
\begin{align}
J\left(t\right)&=  -i a t_{0} R\left(\psi\right)\left({\rm e}^{-i\left[\Phi\left(t\right)-\theta\left(t\right)\right]}-{\rm e}^{i\left[\Phi\left(t\right)-\theta\left(\psi\right)\right]}\right)\nonumber \\
&=  -2 a t_{0} R \left(\psi\right)\sin(\Phi\left(t\right)-\theta\left(\psi\right)),\label{eq:currentexpectation} \\
&\implies \Phi(t)=\theta(\psi)-\arcsin\left(\frac{J(t)}{2at_0R(\psi)}\right).
\end{align}
If this expression is then substituted back into Eq.\eqref{eq:Hamiltonian}, we obtain
\begin{align}
\hat{H}_{J}\left(J(t), \psi\right) & = \sum_{\sigma,j} \left[ P{\rm e}^{-i\theta\left(\psi \right)}\hat{c}_{j\sigma}^{\dagger}\hat{c}_{j+1\sigma} + \mbox{h.c.} \right] +\hat{U}, \label{eq:trackingHamiltonian} \\
P & = -t_{0}\left(\sqrt{1 - X^2(t,\psi)}+i X(t,\psi)\right), \\
   X(t,\psi) & = \frac{J\left(t\right)}{2at_{0}R\left(\psi\right)}. 
\end{align}

By parametrising the Hamiltonian in terms of an expectation, it becomes dependent on $\psi$, and is therefore explicitly non-linear. The equation of motion generated by Eq.\eqref{eq:trackingHamiltonian} is guaranteed to produce the same response $J(t)$ as evolving using Eq.\eqref{eq:Hamiltonian}, but its nonlinear nature means that the uniqueness of evolutions using $\hat{H}_J$ is not assured.  In fact, the evolution is only provably unique if $X(t,\psi)$ satisfies certain conditions, which can be ascertained by considering the term $P$. 

If $P$ is an LC function of $X(t,\psi)$, then it is possible to show $\hat{H}_J$ is \emph{also} LC, and hence generates a unique equation of motion. Note however the converse is also true - if $P$ is not LC, then neither is $\hat{H}_J$. Hermiticity guarantees $X(t,\psi) \leq 1$, but $P$ is only LC if $X(t,\psi) < 1$. This latter condition is due to the fact that $f(x)=\sqrt{1-x^2}$ (and hence $P$) is only LC on the interval $I=[-(1-\epsilon),1-\epsilon]$ (where $\epsilon$ is some finite constant). This ultimately implies that if at any point in the evolution, the condition
\begin{equation}
    \label{eq:uniquenesscondition}
    \left|J(t)\right|= 2 at_0R(\psi) \implies \Phi(t)-\theta(\psi)=n\pi +\frac{\pi}{2},   n \in \mathbb{Z},
\end{equation}
is fulfilled, then the evolution is no longer guaranteed to be unique. 

\begin{figure}
    \centering
    \includegraphics[width=1.0\linewidth]{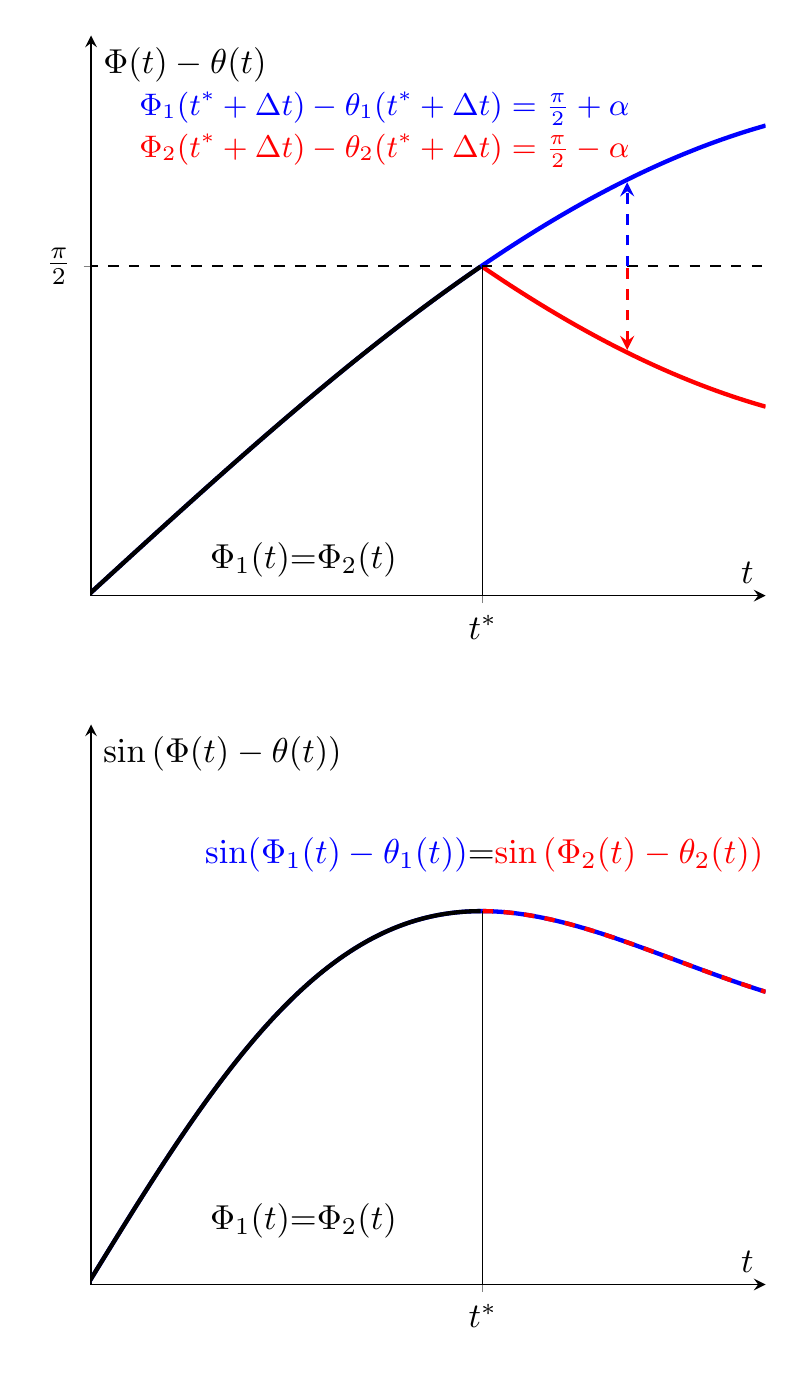}
    \caption{For a trajectory passing through $\Phi(t)-\theta(t)=n\pi +\frac{\pi}{2}$, there are two continuous solutions which at later times which will generate the same $\sin\left(\Phi(t)-\theta(t)\right)$, and hence the same $J(t)$. Importantly however, the Hamiltonian depends not just on the sine but the cosine, meaning that the two solutions which produce the same current do \emph{not} have the same dynamics. }
    \label{fig:splitting}
\end{figure}

Of course, not being provably unique is distinct from being provably non-unique, but this latter condition may be demonstrated by consideration of both $J(t)$ and $\hat{H}_J$ around the points in time where Eq.\eqref{eq:uniquenesscondition} is satisfied. Let us consider two identical systems $\psi_1$, $\psi_2$ driven by fields $\Phi_1(t)$ and $\Phi_2(t)$. For $J_1(t)=J_2(t)$, we require $\Phi_1(t)=\Phi_2(t)$  up to the time $t^*$, when Eq.\eqref{eq:uniquenesscondition} is fulfilled. At this point (taking $n$=0 for simplicity) , we have:
\begin{equation}
    \Phi_1(t^*)-\theta_1(t^*)=\Phi_2(t^*)-\theta_2(t^*)=\frac{\pi}{2}
\end{equation}
where the time dependence of the system is made explicit using $\theta_n(t)=\theta(\psi_n(t))$. Let us then say that a short time later at $t^*+\Delta t$, we have 
\begin{equation}
   \Phi_1(t^*+\Delta t)-\theta_1(t^*+\Delta t)=\frac{\pi}{2}+\alpha,
\end{equation} such that
\begin{equation}
J_1(t^*+\Delta t)=2at_0R(t^*+\Delta t) \cos(\alpha). 
\end{equation}
The same current can be achieved for the second system if it satisfies $\Phi_2(t^*+\Delta t)-\theta_2(t^* +\Delta t)=\frac{\pi}{2}-\alpha$, as illustrated in Fig. \ref{fig:splitting}. 

\begin{figure}
    \centering
    \includegraphics[width=1.0\linewidth]{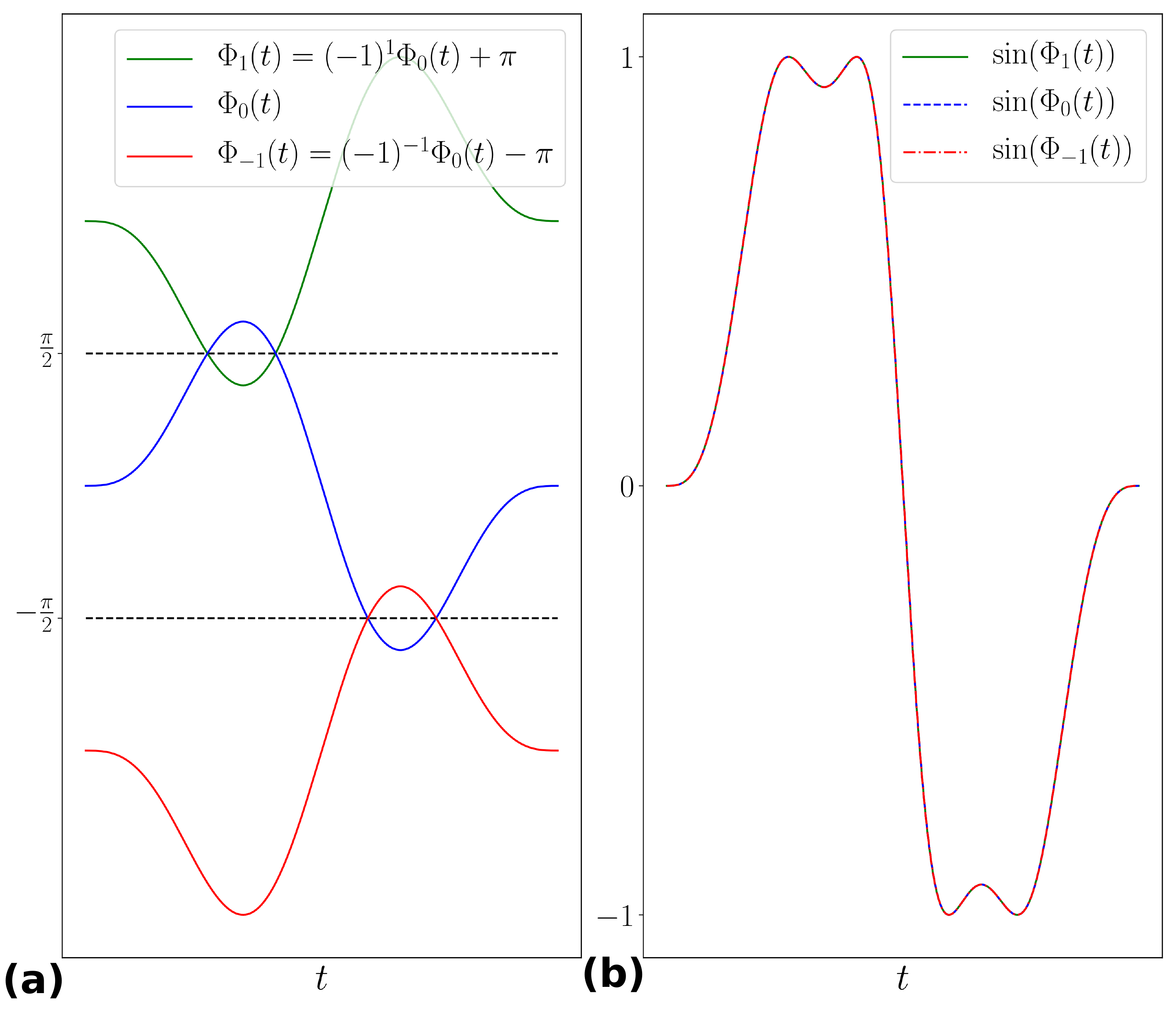}
    \caption{$\Phi(t)$ (suppressing $\theta(\psi)$ for simplicity) has multiple solutions depending on the branch of $\arcsin_l$.  Three possible $\Phi(t)$ are shown in \textbf{(a)}, which all reproduce the same current in \textbf{(b)}. These three trajectories correspond to $\Phi_0(t)$, the principal $\arcsin_0$ solution, and $\Phi_{\pm1}$, which are obtained via the $\arcsin_{\pm1}$ solutions. The critical points described by Eq.\eqref{eq:uniquenesscondition} correspond to the branch points where it is possible to switch continuously from one solution to another.}
    \label{fig:arcsines}
\end{figure}

These two solutions' non-uniqueness are ultimately due to the multiplicity of the $\arcsin(x)$ function, which defines $\Phi(t)-\theta(\psi)$ in relation to $J(t)$ via Eq.\eqref{eq:currentexpectation}. The principal branch $\arcsin_0$ can be related to the branch $\arcsin_l$ via 
\begin{equation}
    \arcsin_l(x)=(-1)^l\arcsin_0(x) +l\pi.
\end{equation}
From this perspective, the points at which the evolution is not LC and non-uniqueness can occur are those branch points where two branches of $\arcsin$ intersect. In this case the evolution of $\Phi(t)-\theta(\psi)$ can switch branch without a discontinuity, which is shown in Fig. \ref{fig:arcsines}. 
 
 Finally, we emphasise that while the branch points in $\arcsin$ allow for identical currents, the full Hamiltonian (and therefore dynamics) relies not only on $\sin(\Phi(t)-\theta(\psi))$, but $\cos(\Phi(t)-\theta(\psi))$. This means that two evolutions satisfying $J_1(t)=J_2(t)$ will evolve according to distinct dynamics at $t>t^*$, such that system expectations other than $J(t)$ will no longer be identical.

\section{Numerical Illustrations \label{sec:Numerics}}
Here we numerically demonstrate the existence of response non-uniqueness, and the physical consequences of a particular choice of $\Phi(t)$. For concreteness, we take $\hat{U}=U\sum_j \hat{c}^\dagger_{j \uparrow}\hat{c}_{j \uparrow} \hat{c}^\dagger_{j \downarrow}\hat{c}_{j \downarrow}$, i.e. the Fermi-Hubbard model. Despite the simplicity of this model, its dynamics display a number of interesting features  \cite{Tasaki1998, fabianessler2005,PhysRevResearch.3.013017}. The most relevant to the current discussion is that, depending on the value $U$, this model displays a highly non-linear response to driving \citep{Ghimire2012, Ghimire2011a, Murakami2018,Silva2018,mccaul_driven_2020,Murakami2021}, which is critical for non-trivial non-uniqueness. 

In all cases, we use an $L=10$ site system at half filling ($n_\uparrow =n_\downarrow =5$), with a hopping parameter $t_0= 0.52$eV. The initial driving field from which other solutions will be derived is 
\begin{equation}
\label{eq:refphi}
    \Phi(t)=A \sin^2\left( \frac{\omega_0 t}{2T}\right)\sin(\omega_0t).
\end{equation}
Where $\omega_0 = 32.7$THz, and the pulse has a duration of $T=2$ periods. The prefactor $A =\frac{aF_0}{\omega_0}$ is constituted from the lattice constant $a=4\AA$ and electric field amplitude $F_0=10$MV/cm, while all figure units are reported in a$^\prime$.u. (atomic units scaled to $t_0$).

Calculations in this section are implemented via the \emph{QuSpin} \cite{weinberg_quspin_2017,weinberg_quspin_2019} package in Python, evolving the system from the ground state via exact diagonalisation. Additional solutions are then generated using \emph{SCOOP} \cite{hold-geoffroy_once_2014} to spawn a suitably modified concurrent process whenever Eq.\eqref{eq:uniquenesscondition} is satisfied. Finally, it should be mentioned that explicit integration methods will converge very poorly at the points where LC is violated \cite{suli_introduction_2003}, and we therefore employ an implicit method to evolve the system.

\subsection{$U=0$ case}
A particularly instructive (and analytically tractable) demonstration of non-uniqueness is in the case of the free particle, where $\hat{U}=0$. In this instance, Eq.\eqref{eq:Hamiltonian} is diagonalisable, using the transformation $\hat{c}_j=\sum^{k=L}_{k=1} {\rm e}^{i\omega_k}\tilde{c}_{k}$, where $\omega_k = \frac{2\pi k}{L}$. The diagonalised Hamiltonian is then:
\begin{equation}
\hat{H}\left(t\right)=-2t_{0}\sum_{k,}\cos(\omega_k-\Phi(t))\tilde{c}_{k}^{\dagger}\tilde{c}_{k}.
\end{equation}
where the spin index has been suppressed (as it plays no role in dynamics). By the same token, it is possible to express the current operator as 
\begin{equation}
\hat{J}\left(t\right)=-2 a t_{0}\sum_{k}\sin(\omega_k-\Phi(t))\tilde{c}_{k}^{\dagger}\tilde{c}_{k}.
\end{equation}

\begin{figure}
    \centering
    \includegraphics[width=1.0\linewidth]{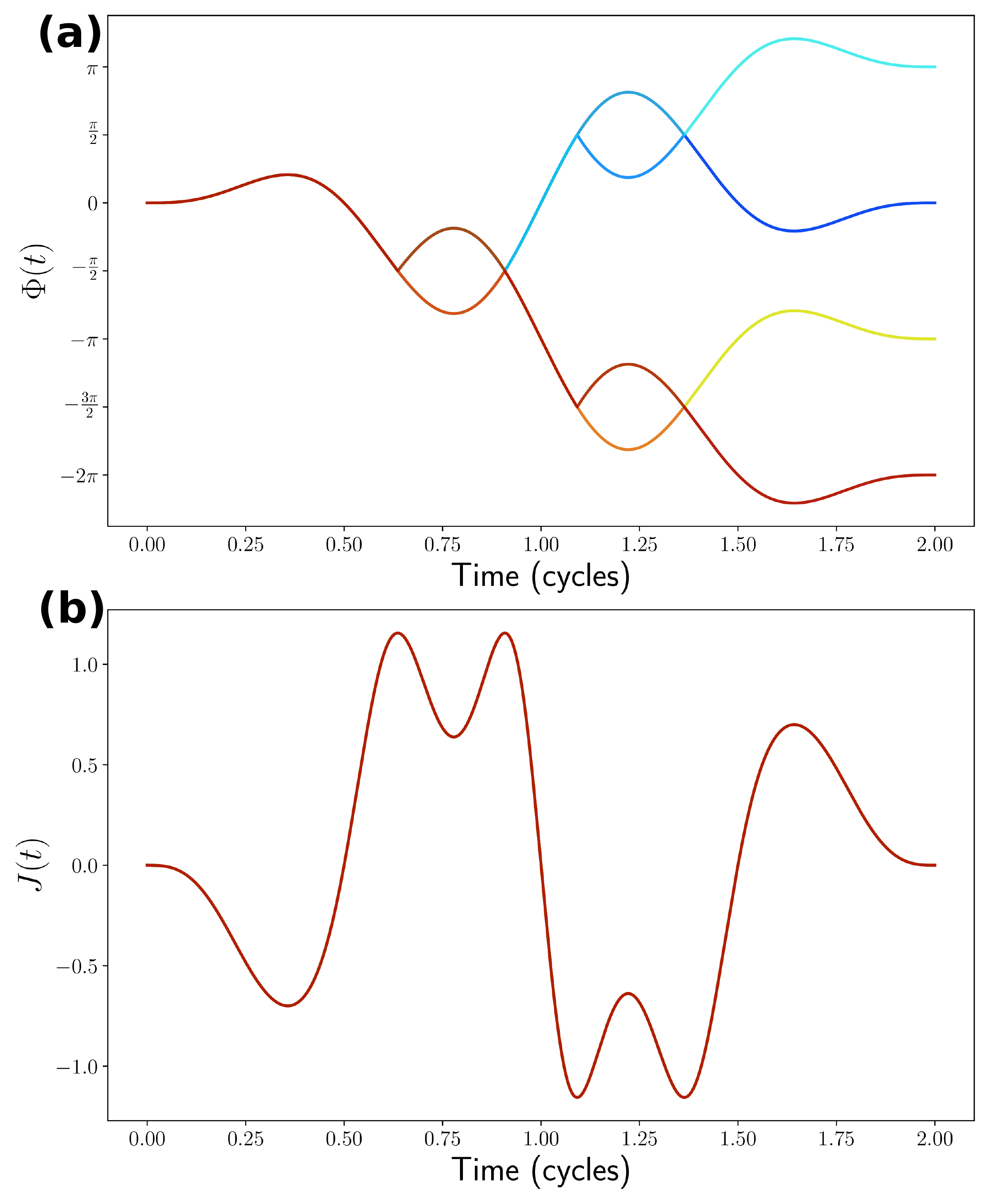}
    \caption{Non-unique optical responses in a free system. \textbf{(a)} All theoretical pulses that will generate the target current \textbf{(b)} The current generated by these pulses. Each color in \textbf{(a,b)} corresponds to different solutions. Not however that many pulses overlap for significant periods, as the only difference between each solution is the choice of branch switch at each critical point.}
    \label{fig:u=0system}
\end{figure}

If one evolves from the ground state, then the set occupied modes will be chosen to maximise $\sum_{k\in \mathcal{K}} \cos(\omega_k)$, where  $\mathcal{K}$ is the set of occupied modes \cite{mccaul_controlling_2020}. One consequence of this at half filling is that 
\begin{equation}
\label{eq:sincancel}
\sum_{k\in\mathcal{K}} \sin(\omega_k)=0  \implies \theta(\psi)=0.
\end{equation}
Even in the case that the system evolves from an excited state or different filling fraction, the nearest neighbour operator and Hamiltonian share an eigenbasis, and hence $\theta(\psi)$ will be a constant. This has the important consequence that the degree of non-uniqueness present in the current response is determined purely by the trajectory of $\Phi(t)$.

For any given $\Phi(t)$ that satisfies Eq.\eqref{eq:uniquenesscondition} $m$ times over the course of the evolution, there will be $2^m$ fields which all produce the same current $J(t)$. For the the reference field given in Eq.\eqref{eq:refphi}, $m=4$. The 16 possible driving fields are shown in Fig.\ref{fig:u=0system}, together with confirmation that each elicits an identical response from the free system.   

\begin{figure}
    \centering
    \includegraphics[width=1.0\linewidth]{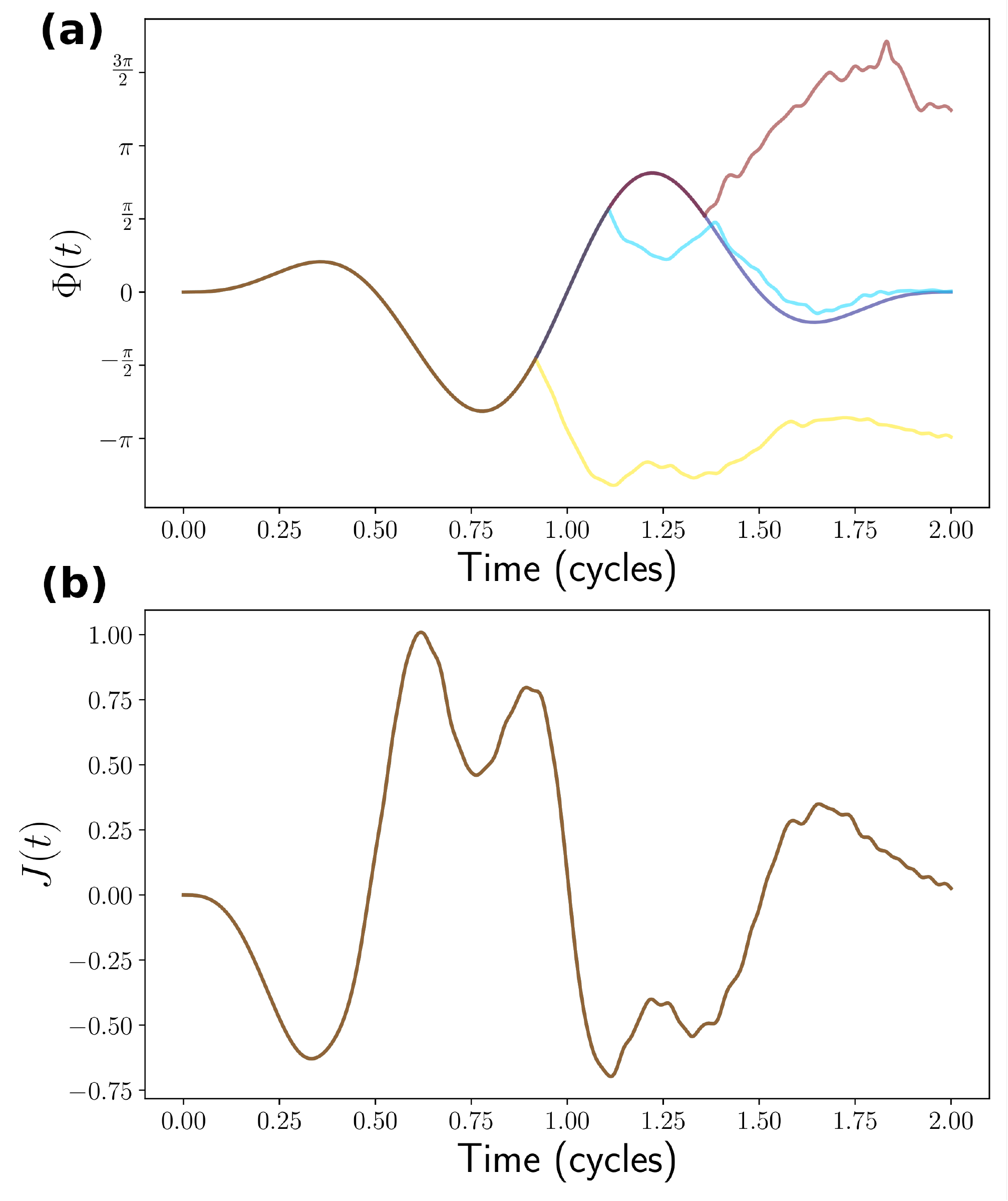}
    \caption{Non-unique response in a $U=0.5 t_0$ Fermi-Hubbard system. \textbf{(a)} shows  a set of pulses found to reproduce the target current, and \textbf{(b)} displays the corresponding currents. Note that some solutions (most notably those generated around the first branch point) are excluded due to numerical instabilities at later times. }
    \label{fig:fermihubbardnonuniqueness}
\end{figure}

\begin{figure}
    \centering
    \includegraphics[width=1.0\linewidth]{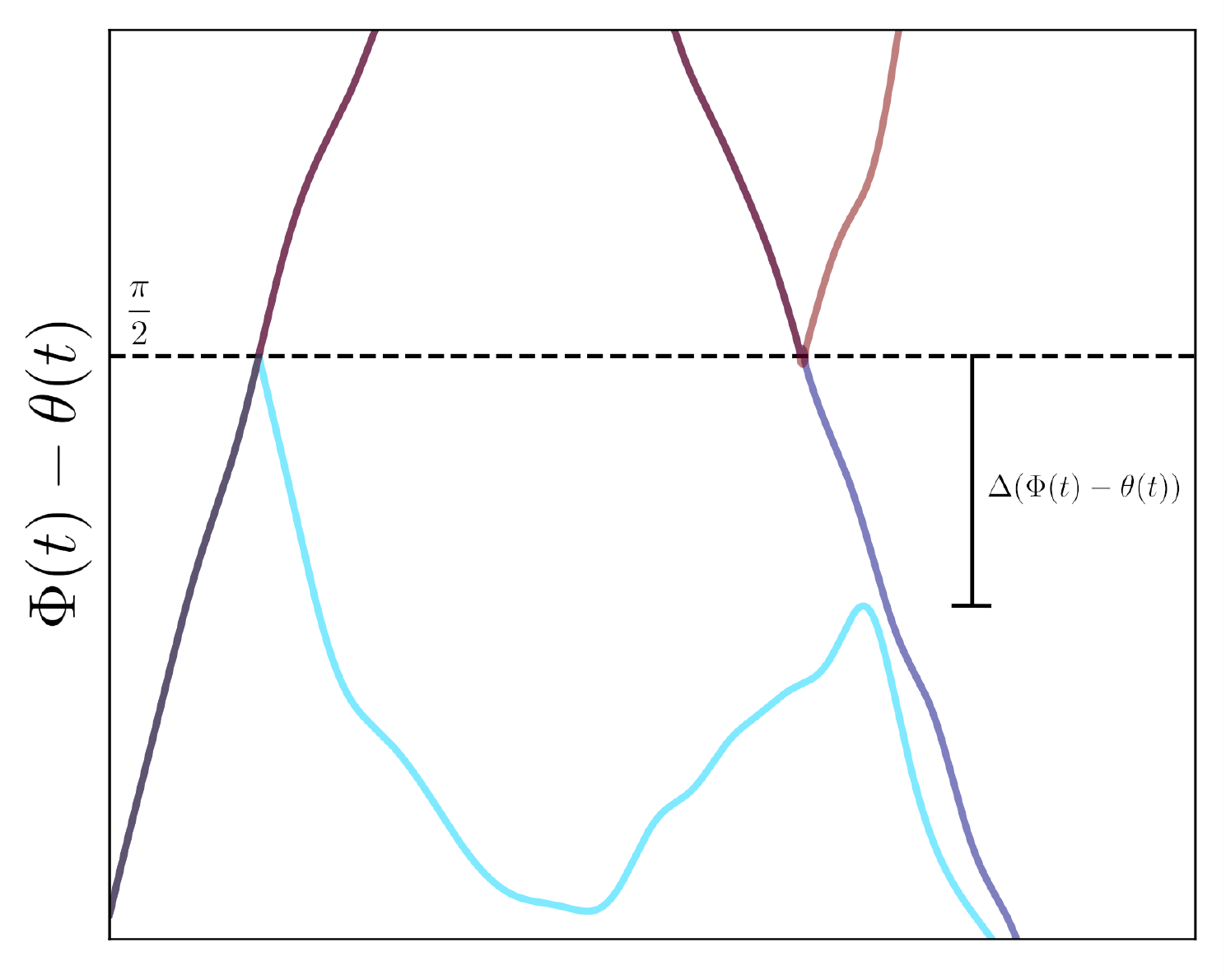}
    \caption{$\Phi(t) - \theta(t)$ in the $U=0.5 t_0$ Fermi-Hubbard system, corresponding to the region around $\sim 1.25$ cycles in Fig.\ref{fig:fermihubbardnonuniqueness}. When any solution has a value of $(2n+1)\frac{\pi}{2}$, it branches into two distinct solutions. When the cyan curve in Fig.\ref{fig:fermihubbardnonuniqueness} hits $\frac{\pi}{2}$, it does not produce a new solution. As evident here, this is due to $\Phi(t) - \theta(t)$ avoiding the branch point. This illustrates the unpredictability of the system dynamics being driven by a new pulse solution.}
    \label{fig:pulsegap}
\end{figure}

\subsection{$U\neq 0$ case}
When $U\neq 0$, $\theta(\psi)$ is no longer a constant of the evolution. This has two important consequences, illustrated in Fig. \ref{fig:fermihubbardnonuniqueness}. The first of these is that while immediately after branching, two solutions may have nearly identical expectations (i.e. $\theta_1(t^* +\Delta t)\approx \theta_2(t^* + \Delta t)$), the fact that they are now evolving under distinct dynamics causes these expectations to diverge. This difference that accumulates over time means that the two solutions are no longer branch switched versions of each other. This then leads to another distinction from the free case - namely that it is impossible to know how many branch points any additional solutions will cross through before calculation. 

An example of this can be seen in the cyan trajectory in Fig.\ref{fig:fermihubbardnonuniqueness} - based on the free system results, one might expect that $\Phi(t)-\theta(\psi)$ would encounter the $\frac{\pi}{2}$ branch point and generate an additional solution at $t\approx \frac{2.8 \pi}{\omega_0}$. In fact, as Fig. \ref{fig:pulsegap} shows, while $\Phi(t)$ crosses the branch point, $\Phi(t)-\theta(\psi)$ avoids it due to the constraint on reproducing $J(t)$. This behaviour demonstrates that in general the degree of non-uniqueness in a response cannot be ascertained beforehand, and must be enumerated via numerical simulation. Nevertheless, one may place a lower bound on this number. If the original $\Phi(t)$ satisfies Eq.\eqref{eq:uniquenesscondition} $m$ times, then there must be at least an additional $m$ solutions, even if none of these cross any further branch points.  

\begin{figure*}[t!]
    \centering
    \includegraphics[width=1.0\linewidth]{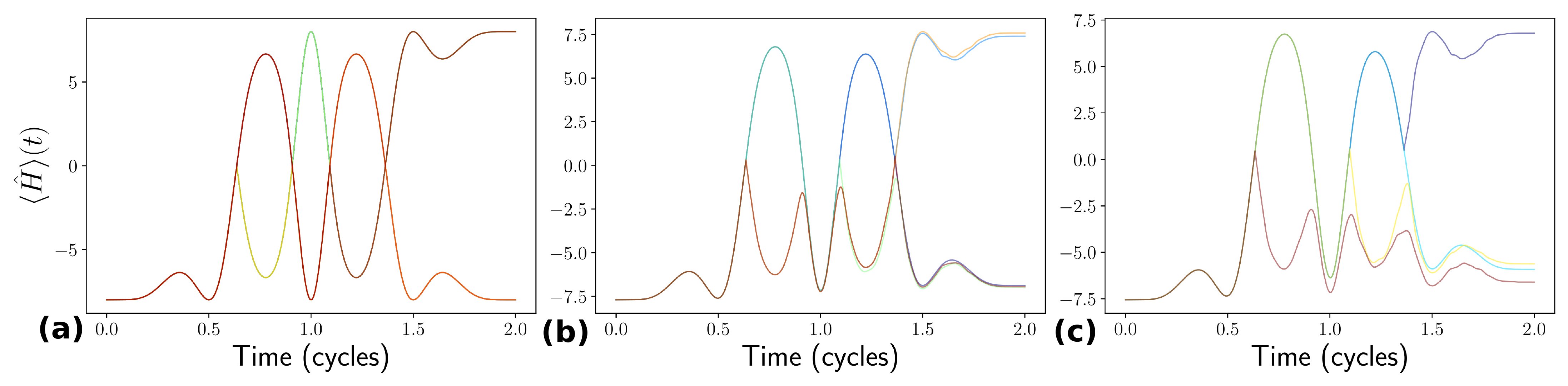}
    \caption{Energy of a Fermi-Hubbard system at half-filling driven from the ground state, where each trajectory corresponds to a distinct driving field $\Phi(t)$, all of which produce the same current. \textbf{(a)} A $U=0$ Fermi-Hubbard system corresponding to a non-interacting system. \textbf{(b)} A $U=0.1 t_0$ system. \textbf{(c)} A $U=0.3 t_0$ system.}
    \label{fig:energy}
\end{figure*}

\subsection{Work done by driving fields}
Having established the multiplicity of driving fields that will produce an identical optical response, we now turn our attention to physical consequences of a particular choice of $\Phi(t)$. Fig. \ref{fig:energy} shows example energetic trajectories for several values of $U$. One of the most striking features is that the work done (defining work as the difference in the Hamiltonian expectation at the final and initial times) by each $\Phi(t)$ is bimodally distributed. Physically, the work done indicates the energy absorbed by the system from the system, and will therefore correspond to modifying the absorption spectrum of the material.

In the $U=0$ free case, this behaviour can again be explained analytically. Consider that if a branch switch occurs such that $\Phi(t)\to -\Phi(t)\pm\pi$, the Hamiltonian expectation will also change:
\begin{align}
\langle\hat{H}(t)\rangle=\cos(\Phi(t))\sum_{k\in\mathcal{K}}\cos(\omega_k)\to- & \cos(\Phi(t))\sum_{k\in\mathcal{K}}\cos(\omega_k) \notag \\ =&-\langle\hat{H}(t)\rangle    
\end{align}
where in the first equality we have exploited the fact that when evolving from the ground state Eq.\eqref{eq:sincancel} applies. Interestingly, if $\langle\hat{H}(t)\rangle$ is the energy of the ground state, then $-\langle\hat{H}(t)\rangle$ will be the energy of the maximally excited state. A driving field containing a branch switch at $t^*$ is therefore dynamically equivalent to evolving with the principal $\Phi(t)$ and maximally exciting the system at $t^*$. The ultimate consequence of this is that the final work done by a given driving field will depend on the number $m$ of branch switches it contains relative to the principal solution. If $m$ is even then $W=0$, but if $m$ is odd then the work done by the  field on the system is $W=2\langle\hat{H}(0)\rangle$. 

In the general case where $U\neq0$, there is again a bimodal splitting in the final system energies, but the presence of interactions lifts the degeneracy in work done between different $\Phi(t)$. This becomes more pronounced as the interaction strength is increased, and the principal field itself does more work on the system.

\section{Discussion \label{sec:Discussion}}
 In this paper we have examined the relationship between a driving field and the response it causes in a quantum system. One might expect that due to the linearity of the Schr{\"o}dinger equation, there is a unique field which generates a given response. Here however he have shown that contrary to this expectation, there are a multiplicity of driving fields which will all generate the same response. To paraphrase Chuck Palahniuk, a system's response to driving is not a `beautiful and unique snowflake' \cite{palahniuk1996fight}.
 
 Using techniques from tracking control, it is possible to identify branch points in the a driving field $\Phi(t)$, from which additional solutions can be generated. These are guaranteed to produce the same response $J(t)$, but have dynamics distinct from each other. One consequence of this is that different solutions for $\Phi(t)$  will do different amounts of work on the system (which in turn affects absorption spectra), illustrating secondary effects that depend on the choice of $\Phi(t)$.  While in the present work we have considered only the current expectation, it is possible to derive tracking control Hamiltonians for an arbitrary expectation \cite{mccaul_controlling_2020}, and hence the same analysis can be extended to derive uniqueness conditions for any given observable.
 
 Finally, it is worth pausing to consider the physical implications and applications of response non-uniqueness. The fact that it is enabled only by the original response crossing critical values means that the degree of non-uniqueness will depend greatly on the response being considered. Nevertheless, non-uniqueness offers some freedom in setting the system state post-driving, which has the potential to be exploited in (for example) a quantum heat engine \cite{doi:10.1146/annurev-physchem-040513-103724}. This phenomenon may also enable the implementation of memristor-like elements in optical computing \cite{https://doi.org/10.1002/adfm.202005582}, where the internal system state can be used as a store of memory of previous pulses. 
 
 More broadly, non-uniqueness can be placed within a family of recent results regarding driven quantum systems - namely that under quite general conditions they are universal (one can always find a driving field to produce an arbitrary response), and that there exists a `twinning field' for any pair of distinct systems, which will elicit the same response from each of them \cite{mccaul_indistinguishability}. In this context, the existence of non-uniqueness provides further evidence of the fundamental malleability of driven systems. That is, not only can one engineer almost arbitrary expectations in a quantum system, there is a degree of choice in how to achieve this.   
\begin{acknowledgments}
 This work has been generously supported by Army Research Office (ARO) (grant W911NF-19-1-0377; program manager Dr.~James Joseph). The views and conclusions contained in this document are those of the authors and should not be interpreted as representing the official policies, either expressed or implied, of ARO or the U.S. Government. The U.S. Government is authorized to reproduce and distribute reprints for Government purposes notwithstanding any copyright notation herein. This work has also been partially supported by by the W. M. Keck Foundation. 
\end{acknowledgments}

\section*{Author Contributions}
{A.K. and G.M. contributed equally to this work}. 
\providecommand{\noopsort}[1]{}\providecommand{\singleletter}[1]{#1}%

\end{document}